# Online Monitoring software framework in the ATLAS experiment

M. Barczyk, D. Burckhart-Chromek[1], M. Caprini[2], J. Da Silva Conceicao, M. Dobson, J. Flammer, R. Jones, A. Kazarov[3], S. Kolos[3,4], D. Liko, L. Lucio, L. Mapelli, I. Soloviev[3]
*CERN, Geneva, Switzerland*

R. Hart
*NIKHEF, Amsterdam, Nederland*

A. Amorim, D. Klose, J. Lima, L. Pedro
*FCUL, Lisbon, Portugal*

H. Wolters
*UCP, Figueira da Foz, Portugal*

E. Badescu
*NIPNE, Bucharest, Romania*

I. Alexandrov, V. Kotov, M. Mineev
*JINR, Dubna, Russia*

Yu. Ryabov
*PNPI, Gatchina, St. Petersburg, Russia*

[1] presenter at the conference
[2] on leave from NIPNE, Bucharest, Romania
[3] on leave from PNPI, St. Petersburg, Russia
[4] paper editor

A fast, efficient and comprehensive monitoring system is a vital part of any HEP experiment. This paper describes the software framework that will be used during ATLAS data taking to monitor the state of the data acquisition and the quality of physics data in the experiment. The framework has been implemented by the Online Software group of the ATLAS Trigger&Data Acquisition (TDAQ) project and has already been used for several years in the ATLAS test beams at CERN. The inter-process communication in the framework is implemented via CORBA, which provides portability between different operating systems and programming languages. This paper will describe the design and the most important aspects of the online monitoring framework implementation. It will also show some test results, which indicate the performance and scalability of the current implementation.

## 1. INTRODUCTION

ATLAS [1] is one of the four experiments in the Large Hadron Collider (LHC) [2] accelerator at CERN. The ATLAS detector consists of several sub-detectors, which in turn are subdivided into a number of partitions, which can be operated in parallel and fully independently one from another.

The data rate from the whole detector after Level 1 Trigger rejection is about 150 Gbyte per second. These data are spread amongst 1600 read-out links, with each of them running at a possible maximum rate of 160 Mbyte per second. The ATLAS Trigger and Data Acquisition (TDAQ) [3] system consists of the High Level Trigger (HLT), which performs event selection reducing data by a factor of 300, and the Data Acquisition system (DAQ), which transports event data from the detector readout to the HLT system and selected events to mass storage.

In order to provide the required functionality and to handle the physics data rate, the TDAQ system will use several thousand processors connected altogether over a high-speed network with each of them running several TDAQ software applications. Monitoring of such a large and complicated system is a vital task during the data taking periods as well as during the commissioning of the detector.

## 2. THE ONLINE SOFTWARE

The Online Software [4] is a sub-system of the TDAQ, which encompasses the software to configure, control and monitor the TDAQ and detectors. It is a customizable framework, which provides essentially the 'glue' that holds the various sub-systems together. It does not contain any elements that are detector specific as it is used by all the various configurations of the TDAQ and detector instrumentation.

The Online Software consists of three main parts responsible for a clearly defined functional aspect of the whole system:

- Control framework - supports TDAQ system initialization and shutdown, provides control command distribution and synchronization, error handling and system verification.
- Databases framework – responsible for configuration of the TDAQ system and detectors.
- Monitoring framework - provides software for the TDAQ system and detector monitoring.

## 3. MONITORING FRAMEWORK

There are many essential parameters in the TDAQ system, which must be continuously monitored during data taking: physics data quality and integrity, consistency of the trigger information, correlation between sub-detectors, status of the hardware and software system elements, etc. This information can be taken from different places in the data flow chain between detectors and the mass storage.

Another important aspect of the monitoring is error reporting. Any malfunctioning part of the experiment must be identified and signaled as soon as possible so that it can be cured.

### 3.1. Monitoring framework model

In the large highly distributed system it must be possible to transport the monitoring information from the places where it is produced to the places where it can be processed. The Monitoring framework, provided by the Online Software, performs this task as it is shown in Figure 1. In this figure the application which produce the monitoring information are called the Monitoring Providers, and the Monitoring Consumers are the applications, which can process this information.

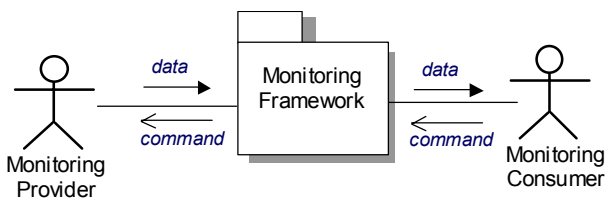

Figure 1: Monitoring framework model.

In addition to the transportation of the monitoring data, the Monitoring framework provides a possibility to transport the monitoring data requests (commands) from the Consumers to Providers.

### 3.2. Monitoring data types

There are different types of information, which can be used to understand the state and correct functioning of the TDAQ system and detector. This can be events or event fragments sampled from well-defined points in the data flow chain, various status and statistics information, which reflect the operation of the hardware elements and software processes in the system, and errors which can be detected at different levels of the system. These types are significantly different in terms of data size, update frequency, type of access, number of Providers and Consumers, etc.

Table 1: Monitoring data types

| Type | Format | Production | Access |
|------|--------|------------|--------|
| Samples of physics events | Vector of 4-byte integers | On request | On request |
| Errors | ID + Severity + Text | In case of faults | Via subscription |
| Histograms | Standard histogram formats | Periodically or whenever it is changed | On request and via subscription |
| Other information | User-defined | Periodically or whenever it is changed | On request and via subscription |

Table 1 shows the main monitoring data types along with their most important characteristics.

### 3.3. Monitoring Architecture

In order to optimize the performance of the monitoring in a large and highly distributed TDAQ system a separate service for each major class of the monitoring information is provided. Each service offers the most appropriate and efficient functionality for given information type and provides specific interfaces for both Monitoring Providers and Consumers. Figure 2 shows the architecture of the Monitoring framework.

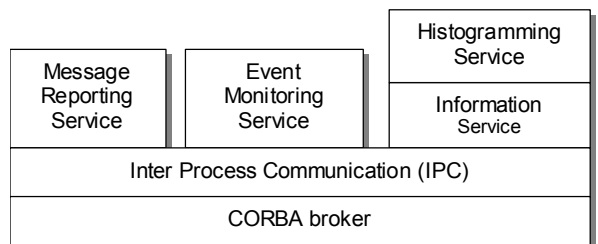

Figure 2: Monitoring framework architecture

The Inter Process Communication (IPC) [5] is a basic communication service, which is common for all the Online Software services. It defines a high-level API for the distributed object implementation and for remote object location. The IPC provides a common basic set of remote methods for all the remote objects in the Online Software. In addition the IPC implements partitioning, which allows to run several instances of the Online Software services in different detector partitions concurrently and fully independently.

The IPC itself is built on top of the Common Object Request Broker Architecture (CORBA) [6] broker, which provides the actual inter-object communication. CORBA is a vendor-independent industry standard defined by the Object Management Group (OMG) [7] for an architecture and infrastructure that computer applications use to work together over networks. The most important features of CORBA are: object oriented communication, inter-operability between different programming languages and different operating systems, object location transparency.

### 3.4. Monitoring Services

#### 3.4.1. Event Monitoring Service

The Event Monitoring Service (EMS) is responsible for transportation of physics events or event fragments sampled from well-defined points in the data flow chain to the software applications, which can analyze them in order to monitor the state of the data acquisition and the quality of physics data of the experiment. An event is transported as a sequence of bytes, so the EMS is neutral to the event format. Figure 3 shows the main interfaces provided by the EMS.

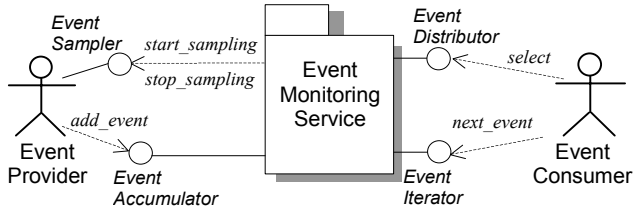

Figure 3: Event Monitoring Service interfaces

The Event Provider which is able to sample events from a certain point of the data flow has to implement the Event Sampler interface. When the Event Consumer requests samples of events from that point via the select method of the *Event Distributor* interface, the EMS system asks this Event Provider to start a sampling process by sending it the *start_sampling* message via the *Event Sampler* interface. The Event Provider samples events and provides them to the EMS via the *Event Accumulator* interface. The Event Consumer can get these events via the *Event Iterator* interface.

When there are no more Event Consumers interested in event samples from a particular point of the data flow chain, the EMS sends the *stop_sampling* message to the appropriate Event Provider via the *Event Sampler* interface to stop the sampling process.

The monitoring framework provides also a graphical user interface application, which is an example of the Event Consumer. The application is written in Java and is called Event Dump. It uses the *Event Distributor* and the *Event Iterator* interfaces to get an event from the place specified by the user and displays the event content.

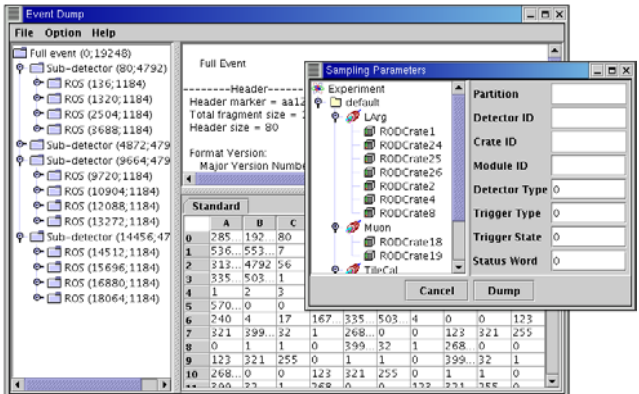

Figure 4: Event Dump application

Figure 4 shows the main window of the Event Dump (the bigger one), which displays the event data, and the selection window (the smaller one) in which the user can define the detector, crate, and module from which the event is to be taken and also can specify some keyword values, which will be used to select the interesting events.

### 3.4.2. Message Reporting Service

The Message Reporting Service (MRS) transports the error messages from the software applications, which detect the errors to the applications, which are responsible for the error handling. The *MRSStream* interface can be used by any application, which wants to report an error as it is shown in Figure 5. In order to receive the error messages an Error Consumer has to subscribe via the *MRSReceiver* interface for the messages it wants to receive. The MRS will forward the appropriate messages to the interested subscribers via the *MRSCallback* interface.

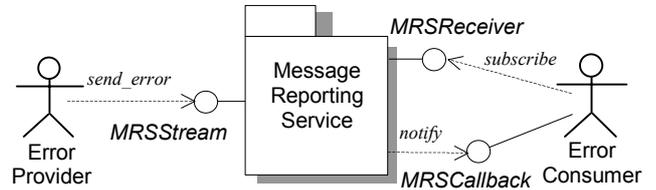

Figure 5: Message Reporting Service interfaces

An example of the Error Consumer is shown in Figure 6. This is the main user control interface application of the TDAQ system, which contains the message display at the bottom.

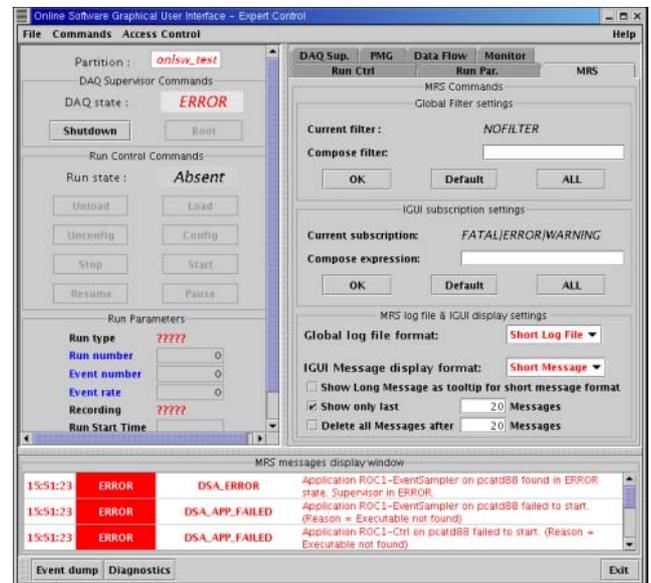

Figure 6: Main TDAQ user interface

This message display shows the error messages coming from the TDAQ applications and detectors. A user can specify the appropriate parameters for the *subscribe* method of the *MRSReceiver* interface via the MRS panel in the right part of the user control window to define the messages to be shown.

### 3.4.3. Information Service

The Information Service (IS) allows TDAQ applications to exchange user-defined information during a run. A user can define the structure of his specific information in XML. Then, he can produce C++ or Java classes using the generator application provided by the IS. The instances of these classes can be shared by the applications. The information structure description is also available at run time.

Figure 7 shows the main interfaces provided by the IS. Any Information Provider can make his own information publicly available by using the *insert* method of the *InfoDictionary* interface and notify the IS about changes of the published information via the *update* method. The *remove* method of the *InfoDictionary* interface can be used to delete the information from the IS.

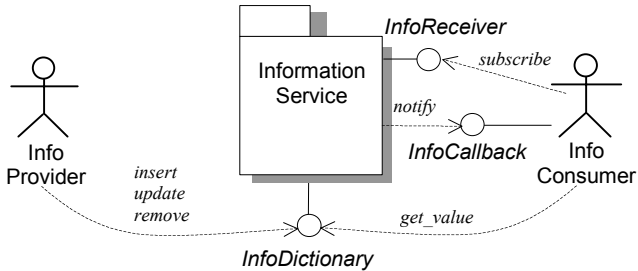

Figure 7: Information Service interfaces

The IS supports two types of information access. It is possible to get the information value directly on request via the *get_value* method of the *InfoDictionary* interface. On the other hand, any Information Consumer can subscribe for a particular information or set of information via the *InfoReceiver* interface, in which case it will be informed about changes of the information for which it subscribed.

There is a graphical user interface which allows to browse the content of the IS. Figure 8 shows the main window of this application (smaller one), which displays all the instances of the IS for a specific detector partition. The partition can be chosen from the list of active partitions in the drop down control at the top-right corner of this window.

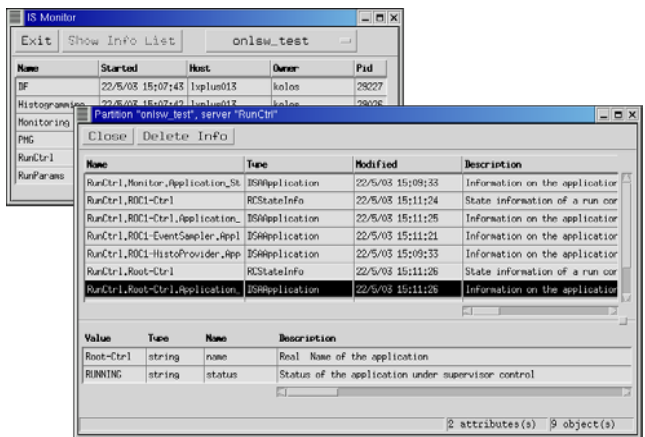

Figure 8: IS Monitor application

Another window in Figure 8 (bigger one) shows the list of information items available at the selected instance of the Information Service. For any item in this list one can see the current value of the information as well as the description of the information item. The content of this window is synchronized with the selected IS instance and is updated automatically whenever the information is changed in the IS.

### 3.4.4. Histogramming Service

The Histogramming Service (HS) allows applications to exchange histograms. From the implementation point of view it is a specialization of the Information Service. The HS defines several information types which are used to transport histograms via the IS.

The HS has an extendable API in terms of the formats of the histograms, which may be used. The HS defines two abstract interfaces: *HistoProvider* and *HistoReceiver*. In order to support a particular histogram format one has to provide an appropriate implementation of those interfaces.

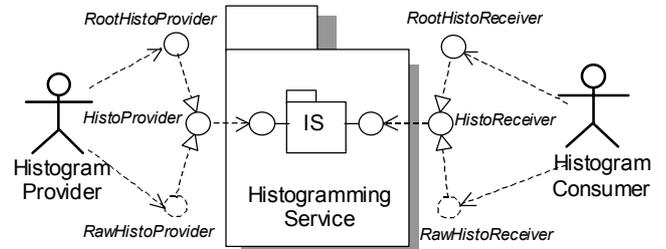

Figure 9: Histogramming Service interfaces

Currently the HS supports two types of histograms as it is shown in Figure 9:
- ROOT histograms – histograms in the format proposed by the ROOT framework [9].
- Raw histograms - histograms represented by arrays of a fundamental data type, i.e. integer, float, double, etc.

Figure 10 shows the Histogram Display application implemented on top of the ROOT Object Browser. This application is an example of the Histogram Consumer, which uses the *RootHistoReceiver* interface to get histograms from the HS.

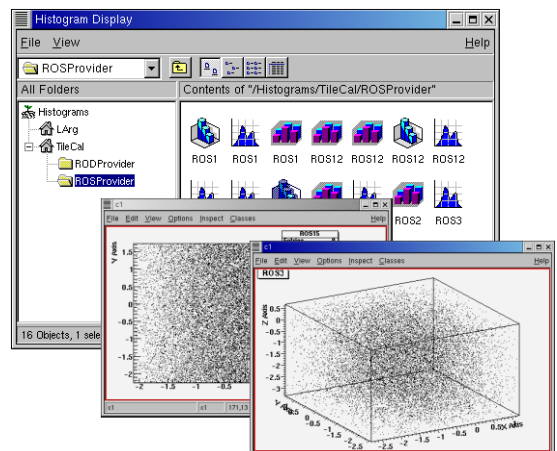

Figure 10: Histogram display

In the left part of the main window (at the background) one can see a list of histogram providers for each detector partition. The right panel shows the list of available histograms for selected provider. Each histogram can be viewed in a separate window, which is an instance of the standard ROOT histogram viewer. This viewer gives

**THGT003**

access to the complete histogram viewing functionality provided by ROOT.

## 4. PROTOTYPE IMPLEMENTATION

Prototype implementations exist for all the Monitoring services. These prototypes are aiming to verify the feasibility of the chosen design and the choice of implementation technology for the final TDAQ system, and are used in the ATLAS test beam operations.

Each service is implemented as a separate software package with both C++ and Java interfaces. All the services are partitionable in the sense that it is possible to have several instances of each service running concurrently and fully independently in different detector partitions.

As it has been mentioned already the services implementation is based on the CORBA. Currently the open source implementation of CORBA provided by Xerox Company is used. It is called Inter Language Unification (ILU) [9]. Several other CORBA implementations are currently being evaluated. They are namely: TAO [10], MICO [11], omniORB [12] and ORBacus [13]. They provide interesting features, which are missing in ILU, i.e. advanced thread management in multy-threaded applications, advanced connection management, CORBA objects persistence, etc. Another CORBA broker can replace ILU without affecting the implementation of the Monitoring services.

### 4.1. Performance and scalability of the current implementation

Among the Monitoring services, the most extensive tests have been performed for the Information Service. The other services are implemented on the same technology and offer the same level of performance and scalability as the IS.

The test bed for the IS tests consisted of 216 dual-Pentium PCs with processor frequency from 600 to 1000 MHz. A single instance of the IS was set up on one dedicated machine. The other 200 machines were used to run from one to five Information Providers on each of them simultaneously. Each Information Provider published one information object in the IS at start up and then updated it once per second. The last 15 machines were used to run 1, 5, 10 or 15 Information Consumers which subscribe for all the information in the IS. Whenever an Information Provider updated his information, this new information was distributed to all the Information Consumers.

Figure 11 shows the average time for transporting information from one Information Provider to all the subscribed Information Consumers as a function of the number of Information Providers working concurrently.

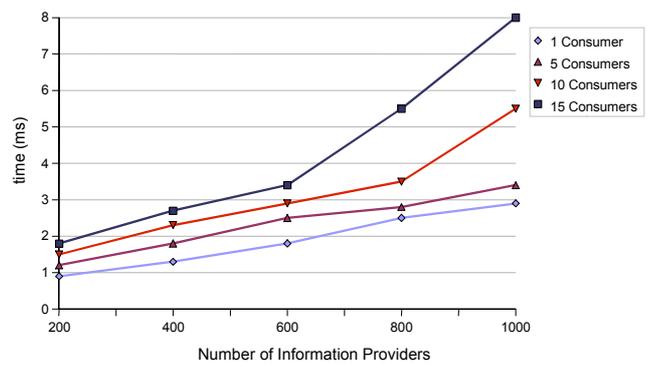

Figure 11: IS test results

The results of the tests show that a single instance of the Information Service is able to handle one thousand Information Providers and about 15 Information Consumers at the same time. In larger configurations the design of the IS allows to distribute the total load among a number of the IS instances, which can run fully independently. Thus, it will be necessary to run only a few (less then 10) IS instances in order to provide the required performance for the final ATLAS TDAQ.

## 5. SUMMARY

Monitoring in the ATLAS experiment is a complex and demanding task. The Online Software of the ATLAS TDAQ system implements a software framework, which can be used for information exchange between the monitoring data providers and consumers. The monitoring framework consists of four services, implemented on top of CORBA. Each service provides the most appropriate and efficient solution for a specific type of the monitoring information. Prototype implementations exist for all the monitoring services and have been successfully used for several years in the ATLAS test beams [4]. The tests, which have been recently performed, show that the services satisfy the requirements of the ATLAS experiment.

**THGT003**